\documentclass[11pt]{article}
\usepackage{geometry}                
\usepackage[parfill]{parskip}    
\usepackage{graphicx}
\usepackage{amsmath, amsthm, amsfonts}
\usepackage{amssymb}
 \usepackage{cite}
\usepackage[width=\textwidth]{caption}

\usepackage{color}
     \usepackage[colorlinks]{hyperref}
\hypersetup{linkcolor=blue,%
citecolor=blue,%
urlcolor=cyan}
\usepackage{url}

\usepackage{epstopdf}
\title{\bf Massive and massless two-dimensional Dirac particles\\ 
in electric quantum dots}

\author{\c{S}. Kuru$^1$\footnote{sengul.kuru@science.ankara.edu.tr, ORCID: \href{http://orcid.org/0000-0001-6380-280X}{0000-0001-6380-280X}}, 
J. Negro$^2$\footnote {jnegro@fta.uva.es, ORCID: \href{http://orcid.org/0000-0002-0847-6420}{0000-0002-0847-6420}}, 
L. M. Nieto$^2$\footnote{luismiguel.nieto.calzada@uva.es, ORCID: \href{http://orcid.org/0000-0002-2849-2647}{0000-0002-2849-2647}}, L. Sourrouille$^3$\footnote {sourrou@df.uba.ar, ORCID: \href{http://orcid.org/0000-0002-8399-6346}{0000-0002-8399-6346}}
\medskip
\\ 
\small
\noindent
$^1$\,Department of Physics, Faculty of Science, Ankara
University, 06100 Ankara, Turkey
\\
\noindent
$^2$\,Departamento de F\'{\i}sica Te\'orica, At\'omica y
\'Optica, and IMUVA,\\ Universidad de Valladolid,  47011 Valladolid, Spain
\\ 
 \noindent
$^3$\,IFISUR-CONICET, Bah\'{\i}a Blanca, Argentina
}

\begin{document}

\maketitle

\begin{abstract}

In this work we investigate the confining properties of charged particles of a Dirac material in the plane subject to an electrostatic potential well, 
that is, in an electric quantum dot. 
Our study focuses on the effect of mass and angular momenta on such confining properties.
To have a global picture of  confinement, both  bound  and  resonance states are considered. 
The resonances will be examined by means of the Wigner time delay of the scattering states, as well as through the complex eigenvalues of outgoing states in order to show that they are physically meaningful.  
By tuning the potential intensity of the well, electron captures and atomic collapses are observed for critical values. 
In these processes, the bound states of the discrete spectrum become resonances of the continuous spectrum or vice versa.
For massive charges, the atomic collapse phenomenon keeps  the number of bound levels in the quantum dot below  a maximum value.
In the massless case, the bound states have zero energy and occur only for some discrete values of the potential depth, as is known. 
We also show that although the intensity of the resonances for massive particles is not significantly influenced by angular momenta, on the contrary, for massless particles they are quite sensitive to angular momenta, as it is the case of graphene.
\end{abstract}



\section{Introduction}

It is well known that electric fields constitute a good tool for confining relativistic massive  charged particles in two or more spatial dimensions (see a detailed discussion by R. Hall et al \cite{hall04} for three dimensions), although  in the case of two-dimensional massless particles they are not so useful due to the strong Klein tunnelling effect \cite{katnelson06,beenakke08}.  
For this reason, to confine massless Dirac electrons in graphene, it is much better to appeal to magnetic fields \cite{castro09,martino07,negro09,negro18,portnoi16b}. 
However, due to its potential importance in technological  applications, the electrical confinement of  massless particles has also been investigated in a recent series of papers \cite{apalkov08,bardarson09,peeters08,Hartmann14,Portnoi17,portnoi11,portnoi12,portnoi19,miserev,ho14}.
Note that since the origin of the transparency of potentials for zero-mass particles in general comes from transverse momenta, electric fields could still be useful, for example, in the study of quantum wires, where the momentum is parallel to the wall of the confining potential. 
Another case in which electric fields could produce confinement is when the quantum dot has a symmetry, for example radial, which allow for a well-defined angular momentum quantum number corresponding to a motion parallel to the quantum dot (there is a close connection between classical non-chaotic trajectories, superintegrability and transparent potentials \cite{bardarson09}).

On the other hand, it is also worth considering massive particles in two dimensions due to the introduction of new materials similar to graphene, such as silicene, germanene, stanene, and phosphorene, quite attractive from the point of view of topological insulators (see  \cite{candido18} and references quoted therein), where the charges acquire an effective mass from the spin-orbit interaction and the perpendicular electric fields \cite{liu11}.
An important consequence of the confinement of massive particles is that it allows to observe the phenomenon of atomic collapse, which is possible but very difficult to experience in relativistic quantum mechanics. 
However, inside Dirac materials  this is more accessible (as well  as other relativistic effects, such as the above mentioned Klein tunnelling) due to the much lower Fermi velocity $v_F$ relative to the speed of light $c$ \cite{peeters17}. 
This effect is shown in experiments that have recently been  designed under different configurations \cite{peeters16, peeters18, peeters19, peeters21}. 
Atomic collapses consist of the decay of states from the discrete spectrum of the quantum dot to the negative energy continuum when the depth of the well (or the charge of the atom) increases sufficiently. 
These relativistic effects  confirm the Dirac behaviour of quasi-particles in two-dimensional Dirac materials.

An important issue closely related to  that of  confinement is the search for resonances,   and indeed, the confinement problem in Dirac materials must necessarily include both  bound and resonance states. In fact, we will show   here how the bound states give rise to resonances at each atomic collapse. 
  However, so far only a   few number of articles have considered resonances or quasi-bound states in this  context. For instance,   the works \cite{apalkov08,peeters08} are very close to our   point of view when trying to describe the relationship between  resonances and bound states in graphene,   although both papers are limited to the massless case, missing the genuine atomic collapse of massive charges. 
Throughout this work, resonances will be reexamined in detail, for example   by showing the wave functions of the critical and supercritical states   that make the transition from resonance to bound states.  
The boundary conditions of the complex eigenvalue problem will also be handled very carefully, as well as the associated phase shift of scattering states   according to  Levinson's theorem \cite{dong98}.

In this context, it is worth to mention refs.~\cite{pereira07,shytov07,gamayun09,gamayun11,peeters18,novikov07} which deal with different aspects of the atomic collapse caused by some types of impurities in graphene,   which essentially consist either in a Coulomb potential, 
the so called Dirac-Coulomb problem, or a constant potential well obtained by a configuration of  gate electrodes,   as in the case of the present article. The minimum charge or coupling constant $g$ which produce atomic collapse, is called `critical coupling', $g_c$, and its potential is   referred to as `critical potential', $V_c$. 
  In this framework, in \cite{pereira07} the local density of states (LDOS) is calculated, which is closely related to resonances, showing the difference when the coupling is greater  or less than the critical value,   while \cite{shytov07} focuses  on the polarization produced by the Dirac-Coulomb impurity and the drastic difference when the coupling constant passes the critical value $g_c$,   the polarization being calculated with the help of the phase shifts of analytic solutions. 
Other works \cite{gamayun09,gamayun11} discuss the
instability due to a Coulomb impurity center and calculate formulas for the critical potential
values where collapses start. In particular, \cite{gamayun11} is a wide work that includes a  study of massive and massless fermions in  cylindrical wells (as in the present work) and in Coulomb potentials too. 
The influence of a constant magnetic field on the atomic collapse is also considered   in \cite{gamayun11} (see also \cite{peeters18}),   where a wide variety of applications related to the critical coupling are given. 
  Remark that in the present work we have focused our attention on very particular aspects: the evolution of bound into resonance states and some properties involved in this process. For example, we will provide a series of graphs that show the behavior of a set of bound states that transform into  resonances as the depth of the well is changing, for gapped and gapless fermions, which give us a global view of charge confinement in a two-dimensional (2D) Dirac material. 
Resonances have been characterized in two ways: by means of outgoing boundary conditions and by the Wigner time delays. The phase shift  for scattering states has been computed showing how they fit to the resonances.  
Finally, in relation to Coulomb impurities, wee should mention  the article \cite{novikov07} where the singularity of the potential of $1/r$ is analyzed and the scattering of gapped and gapless fermions  is studied, applying it to transport in graphene.

From a certain point of view, resonances are characterized by  complex energy values $E= E_R+i E_I$,  where the imaginary part $E_I$ is very small. Then, if the energy of a wave packet incident on the potential well (or barrier) is close from $E_R$, the outgoing wave packet can have a prolonged stay in the region where the potential is significant, giving rise to a quasi-bound state \cite{peeters08}. 
The delay time can be calculated using Wigner's formula for the phase change in  scattering states \cite{wigner,ohanian,gadella17}. 
We should mention that the phase change in a scattering resonance is related to the bound states in each angular momentum channel, due to an extension of Levinson's theorem to the domain of relativistic quantum mechanics on the plane \cite{hall04,dong98}. 
Therefore, in this way the tight link of resonances and bound states is also shown. In fact, when the imaginary part $E_I$ is zero, the value $E_R$ can correspond to a bound state (sometimes it may be associated with what is called antibound state \cite{ohanian}). In the massless case, the bound states are characterized by $E_R=E_I=0$.

The  main objective of this work is to draw a number of conclusions about the role of mass and angular momenta in the problem of confinement in electric dots. 
We will change the intensity in the dot (this can be done by varying the potential of the gates that produce the potential (or using the tip of an STM microscope \cite{peeters16})) and show its effect on resonances and bound states. 
As we will see, resonances manifest a relativistic behaviour, particularly at the critical values of the potential depth they will lead to collapses and the associated scattering phases will satisfy Levinson's theorem near the capture of a bound state.
The most striking difference that we will find between the massive and massless cases is that increasing the depth of the well for massless particles is very sensitive to angular momentum, while for the massive case, the value of the angular momenta affects the intensity of the resonances very slightly. 

The structure of the paper is as follows. Section 2 begins with the study of the bound states, resonance and dispersion of massive particles, showing the effect that increasing the depth of the well has on these states. Section 3 addresses the same problems for massless particles, while the final section is dedicated to presenting the main conclusions.

\section{Bound states and resonances for  massive two-dimensional Dirac particles}

We will start with the 2D Dirac Hamiltonian on a Dirac planar material which describes the interaction
of particles of mass $m$ and charge $e$  with an external electrostatic potential $V({\bf x})$:
\begin{equation}
H = v_F \, {\boldsymbol \sigma}\cdot{\bf p} + m v_F^2\, \sigma_z+  e\, V({\bf x}), 
\qquad  {\bf x}=(x,y)\in\mathbb{R}^2 .
\end{equation}

Here, ${\boldsymbol \sigma} = (\sigma_x,\sigma_y)$ and $\sigma_z$ are Pauli matrices,
${\bf p} = (p_x,p_y) = -i\hbar(\partial_x,\partial_y)$ the momentum operators, and
$v_F$ is the Fermi velocity of the material. To be more specific, we will consider a potential with radial symmetry
$V({\bf x})=V(r)$, so naturally from now on we will use polar coordinates  $(r,\theta)$ to
separate variables in the time-independent Dirac equation. We assume that the dynamics takes place on a  much larger scale than the graphene lattice constant, therefore we will  not consider inter-valley  scattering between the Dirac points and will restrict ourselves to a single Dirac point, $K$ \cite{peeters08,guinea06}.

In the case we are dealing with, the Hamiltonian commutes with the total angular momentum operator defined as
\begin{equation}
J_z = L_z + \Sigma,\quad {\rm with}\quad L_z = -i\hbar \partial_\theta \quad {\rm and}\quad
\Sigma = \frac 12\hbar \sigma_z .
\end{equation}
Thus, we can look for the eigenfunctions $\Phi(r,\theta)$ of $H$ that at the same time are eigenfunctions of $J_z$,
\begin{equation}\label{j}
H \Phi(r,\theta) = E\Phi(r,\theta),\qquad J_z\Phi(r,\theta) = j\hbar \Phi(r,\theta).
\end{equation}
It is quite easy to show that the second of the equations  in (\ref{j}) leads to eigenfunctions  with the following  spinor form
\begin{equation}\label{phi}
\Phi(r,\theta) = \left(\begin{array}{c}
\phi_1(r)e^{i (j-\frac12) \theta}
\\[1.5ex]
i \phi_2(r)e^{i (j+\frac12) \theta}\end{array}\right) , \quad j= \ell +\frac12 , \quad \ell= 0,\pm1,\dots
\end{equation}
where $\ell = j-1/2$ and $\ell + 1 = j + 1/2$ are, respectively, the integer orbital angular momentum of the upper and lower component of the spinor, and the imaginary unit in the second component is introduced for convenience. 

After replacing (\ref{phi}) in the eigenvalue equation for $H$ in (\ref{j}),
we get a reduced equation in the variable $r$:
\begin{eqnarray}\label{spinorialsystem}
\left(\begin{array}{cc}
0 & -i A^-
\\[1.5ex]
iA^+ & 0 \end{array}\right)  
\left(\begin{array}{c}
\phi_1(r)
\\[1.5ex]
i \phi_2(r)\end{array}\right) =
\left(\begin{array}{cc}
E{-}{ e}V(r){-}m v_F^2 & 0
\\[1.5ex]
0 & E{-}{ e}V(r){+}m v_F^2 \end{array}\right)  
\left(\begin{array}{c}
\phi_1(r)
\\[1.5ex]
i \phi_2(r)\end{array}\right),
\end{eqnarray}
where the operators $A^\pm$ are given by
\begin{equation}\label{aminusplus}
A^-= \hbar v_F\left(\partial_r +\frac{\ell+1}{r}\right),\quad  
A^+= \hbar v_F\left(-\partial_r +\frac{\ell}{r}\right),\quad \ell\in \mathbb{Z}.
\end{equation}
In this work we will choose the electric potential to be a typical  two-dimensional radial well of the form
\begin{equation}\label{potentialwell}
V(r) = \left\{\begin{array}{cl}
V_0,  & r<R,
\\[1.5ex]
0,  & r>R.
\end{array}\right.
\end{equation}
We redefine variables in natural units for this problem as
\begin{equation}\label{newvar}
\rho=\frac{r}{R},\quad 
\varepsilon = \frac{E R}{\hbar v_F},\quad 
 v=  \frac{e V_0R}{\hbar v_F}<0 ,\quad 
\mu =  \frac{m v_F R}{\hbar },
\end{equation}
where we will assume that the  effective potential inside the dot ($v$) is constant and negative unless otherwise stated (for $v>0$ we would have the problem of a potential barrier instead of a well), and the potential outside the dot is zero. We could have used another shape for the potential well, but we preferred the option mentioned above to be able to compare the results here obtained with other relevant references available in the literature \cite{apalkov08,bardarson09,peeters08,portnoi19}. 

With the new notation introduced in \eqref{newvar},  equations \eqref{spinorialsystem}-\eqref{aminusplus} become the following coupled differential system
\begin{equation}\label{phis}
\left\{
\begin{array}{l}
\displaystyle 
\phi'_{2,\alpha} (\rho)+\frac{\ell+1}{\rho}\ \phi_{2,\alpha} (\rho)= \varepsilon^{-}_{\alpha} \ \phi_{1,\alpha} (\rho),
\qquad \\[2ex] 
\displaystyle 
-\phi'_{1,\alpha} (\rho)+ \frac{\ell}{\rho}\ \phi_{1,\alpha} (\rho) = \varepsilon^{+}_{\alpha} \  \phi_{2,\alpha} (\rho),
\end{array}\right.
\end{equation}
where the subindex $\alpha$ can be either $\alpha={\rm i}$ in the inner region of the dot ($0\leq\rho<1$), or $\alpha={\rm o}$ in the outer region of the dot ($\rho>1$), being
\begin{equation}\label{varepsilon}
\varepsilon_{\rm i}^\pm=\varepsilon -v \pm\mu,\  0\leq\rho<1 ,
\qquad
\varepsilon_{\rm o}^\pm=\varepsilon \pm\mu,\  \rho>1.
\end{equation}
Next, we will solve this set of equations in these two regions. Note that both the potential $v$  and the energy parameters $\varepsilon^\pm_\alpha$ are constant, although different, in each of these two intervals. 
The connection between the solutions for each of the two regions is obtained by imposing the continuity of the components $\phi_{ 1,\alpha}$ and $\phi_{ 2,\alpha}$ at the point  $\rho=1$. The equation for
$\phi_{ 1,\alpha}$, obtained from (\ref{phis}), is
\begin{equation}\label{phi_1}
\rho^2 \phi''_{ 1,\alpha}(\rho)+\rho\, \phi'_{ 1,\alpha}(\rho)+ (p^2_{\alpha}\, \rho^2 - \ell^2)\phi_{ 1,\alpha} (\rho) =0, 
\end{equation}
where the `momentum' $p$ in each interval $\alpha={\rm i},{\rm o}$ is
\begin{equation}\label{ps}
p_{\rm i}= \sqrt{(\varepsilon-v)^2- \mu^2},\quad  0\leq\rho<1,
\qquad
p_{\rm o}= \sqrt{\varepsilon^2- \mu^2},\quad  \rho>1,
\end{equation}
where $p_{\rm i}$ and $p_{\rm o}$ are valid, respectively, in the inner and outer regions. 
This means that, as long as $p_{\rm i}$ and $p_{\rm o}$ are nonzero (later we will discuss what happens when one of them is zero), the general solution within each interval can be expressed either as a linear combination of Bessel functions of the first and second kind ($J_\ell, Y_\ell$) \cite{abramowitz}, or as a linear combination of Hankel functions of the first and second kind ($H^{(1)}_\ell, H^{(2)}_\ell$), in the form
\begin{equation}\label{phi1}
\phi_{ 1,\alpha}(\rho) = a_{\alpha}\, J_\ell(p_{\alpha} \rho) + b_{\alpha}\, Y_\ell(p_{\alpha} \rho) =\tilde{a}_{\alpha}\, H^{(1)}_\ell(p_{\alpha} \rho) + \tilde{b}_{\alpha}\, H^{(2)}_\ell(p_{\alpha} \rho),
\qquad  \alpha={\rm i},{\rm o}. 
\end{equation}
The arbitrary constant coefficients {$a_{\rm i},b_{\rm i}\dots$  are used for the inner region, while $a_{\rm o},b_{\rm o},\dots$} are used for the outer region. The second  radial function $\phi_{ 2,\alpha}$ of the spinor can be obtained from the previous expression for $\phi_{ 1,\alpha}$ and the second equation of (\ref{phis}).  Using well-known properties of the Bessel and Hankel functions \cite{abramowitz}, we get
\begin{equation}\label{phi2}
\phi_{ 2,\alpha}(\rho) =
\frac{p_{\alpha}}{\varepsilon^{+}_{\alpha}} 
\bigl[ a_{\alpha} \, J_{\ell+1}(p_{\alpha}  \rho) + b_{\alpha} \, Y_{\ell+1}(p_{\alpha} \rho)\bigr]
=
\frac{p_{\alpha}}{\varepsilon^{+}_{\alpha}}   \Bigl[\tilde{a}_{\alpha}\, H^{(1)}_{\ell+1}(p_{\alpha} \rho) + \tilde{b}_{\alpha}\, H^{(2)}_{\ell+1}(p_{\alpha} \rho) \Bigr] ,
\quad   \alpha={\rm i},{\rm o} .
\end{equation}
Now, we are going to use this equation in several situations of physical interest: when there are bound states, resonances or the so-called {\it critical states}. We will assume below that $\ell\geq 0$, since for negative values, although it is not an equivalent situation, the results are similar.

\subsection{Bound states\label{bs}}
To characterize the bound states in the problem we are analyzing, the correct solutions are chosen using the appropriate boundary conditions at the origin $\rho=0$, at the junction point $\rho=1$, and at $\rho\to \infty$.
\begin{itemize}
\item
In the inner region, $0\leq\rho<1$, the Bessel functions that are bounded at the origin
are only those of the first kind (even in the case where $p_{\rm i}$ be a complex number), so, according to \eqref{phi}, in this interval the solutions must have the form
\begin{equation}\label{phileft}
\Phi_{\rm i}(\rho,{\theta}) = a_{\rm i} \left(
\begin{array}{c}
J_\ell(p_{\rm i} \rho)\  e^{i \ell \theta}
\\[1.5ex]
\displaystyle i \frac{p_{\rm i}}{\varepsilon_{\rm i}^+} J_{\ell+1}(p_{\rm i} \rho) \ e^{i (\ell+1) \theta}
\end{array}\right),
\ \ell\in \mathbb{Z}.
\end{equation}
The special value $p_{\rm i}=0 \implies \varepsilon - v=\pm\mu$, gives no additional solution.
\item
In the outer region, $\rho>1$, the appropriate solution to study  bound states is the  Hankel function of the first kind, since its asymptotic behavior when $\rho\to \infty$ is
\begin{equation}\label{asympH1}
H_\ell^{(1)}(p_{\rm o} \rho) \sim \sqrt{\frac{2}{\pi p_{\rm o}  \rho}} 
e^{i(p_{\rm o}  \rho -\ell\pi/2-\pi/4)} .
\end{equation}
\end{itemize}
Therefore, from \eqref{asympH1} it is clear that bound states will appear only if $p_{\rm o}=\sqrt{\varepsilon^2-\mu^2}$ is an imaginary number, $p_{\rm o}=i\, {\rm Im}(p_{\rm o})$, with
 ${\rm Im}(p_{\rm o})>0$ (or possibly zero, see below), that is 
\begin{equation}\label{conditionbound}
\varepsilon^2<\mu^2 \ \implies \ -\mu<\varepsilon<\mu,
\end{equation}
where $\varepsilon$ and $\mu$ are the energy and the mass in the units defined in (\ref{newvar}).
In other words, relativistic bound states in electric fields can only take place for a range of energy which is bounded from below/above by minus/plus the particle mass.
The special cases $\varepsilon = \pm \mu$ where $p_{\rm o}=0$ are {\it called critical points} and will be studied
separately  in the next subsection.
Thus, the wave function of the bound states in the outer region must take the form 
\begin{equation}\label{phiright}
\Phi_{\rm o}(\rho,{\theta}) = \tilde{a}_{\rm o} \left(
\begin{array}{c}
H^{(1)}_\ell(p_{\rm o} \rho)\ e^{i \ell \theta}
\\[1.5ex]
\displaystyle i \frac{p_{\rm o}}{\varepsilon_{\rm o}^+} H^{(1)}_{\ell+1}(p_{\rm o} \rho) \ e^{i (\ell+1) \theta}
\end{array}\right),
\ \ell\in \mathbb{Z}.
\end{equation}
Finally, the eigenvalues of the bound states are obtained by imposing the condition of continuity of the two spinor functions  (\ref{phileft}) and (\ref{phiright})  at the point $\rho=1$, obtaining the following secular equation:
\begin{equation}\label{secular} 
\varepsilon_{\rm i}^+\,  J_\ell(p_{\rm i})\,  p_{\rm o}\, H^{(1)}_{\ell+1}(p_{\rm o})  -
   \varepsilon_{\rm o}^+\,  H^{(1)}_\ell(p_{\rm o})\, p_{\rm i}\, J_{\ell+1}(p_{\rm i})=0.
\end{equation}
Once fixed $v$ and $\mu$, the solutions corresponding to the discrete values of the energy $\varepsilon $ that arise from the secular equation  (\ref{secular}) will have their corresponding eigenfunctions
$\Phi(\rho,\theta) $ constructed from the matching of (\ref{phileft}) and (\ref{phiright}).

\subsection{Critical and supercritical states}

The critical points are the eigenvalues corresponding to bound (or quasi--bound) states such that 
$p_{\rm o}=0$ or $\varepsilon^2=\mu^2$, that is, they correspond to the maximum $\varepsilon= \mu$ or minimum $\varepsilon= -\mu$ possible eigenvalues of the energy (the latter case is often called supercritical). Critical eigenvalues can only be reached for some special values of the potential depth $v$.
The associated eigenstates are called critical and supercritical states. In other words, we must look for the possible values of the potential depth $v$, so that there are states with eigenvalue $\varepsilon=\mu$ or $\varepsilon=-\mu$ and with a bounded behavior such as $\rho\to0$  and  
$\rho\to\infty$, corresponding to a square integrable function (or at least bounded for quasi-bound states).

\subsubsection{Critical states: $\varepsilon= \mu$}
Inside the potential well those states are described by Bessel functions, as in \eqref{phileft}, with the following values of the parameters that appear there: $\varepsilon^{+}_{\rm i} = 2\mu-v>0$ and $p_{\rm i} = \sqrt{v^2-2v\mu}>0$.

In the outer region, taking into account that $\varepsilon_{\rm o}^+ = 2\mu$ and $\varepsilon_{\rm o}^- = 0$, the components $\phi_{1,{\rm o}}(\rho)$ and $\phi_{2,{\rm o}}(\rho)$ satisfy this particular form of equations (\ref{phis})
\begin{equation}\label{phis2}
\left\{
\begin{array}{l}
\displaystyle 
\phi'_{2,{\rm o}} (\rho)+\frac{\ell+1}{\rho}\,\phi_{2,{\rm o}} (\rho) = 0 ,
\\ [2ex]
 \displaystyle -\phi'_{1,{\rm o}} (\rho) + \frac{\ell}{\rho}\,\phi_{1,{\rm o}} (\rho) = 2\mu \phi_{2,{\rm o}} (\rho)\,.
\end{array}
\right.
\end{equation}
Then, $\phi_{1,{\rm o}}$ satisfies an Euler equation (see the limit $p_{\rm o}\to 0$ of (\ref{phi_1}))  whose acceptable solutions give the outer spinor
\begin{equation}\label{phirightc}
\Phi_{{\rm o}}(\rho,\theta) = a_{\rm o} \left(
\begin{array}{c}
\displaystyle \rho^{-\ell} \ e^{i \ell \theta}
\\[1ex]
\displaystyle i  \frac{\ell}{\mu} \,\rho^{-(\ell+1)}\ \ e^{i (\ell +1) \theta}
\end{array}\right),
\ \ell= 1,2,\dots
\end{equation}
Note that in the case $\ell=0$ the critical state is not really a bound state, but simply a {\it quasi-bound state}: the wave function is  not square integrable, although it is bounded. For $\ell=1$ the wave function tends to zero but it is not yet square integrable.
For the following values $\ell=2,\dots$, the critical state wave functions, according to (\ref{phirightc}), satisfy all the conditions to represent true bound states. These types of critical state solutions will also be discussed in detail for the massless case of the next section. 

The matching condition of the solutions of the critical wave functions (\ref{phileft}) and (\ref{phirightc}) at $\rho=1$ produce the following secular equations:
\begin{equation}\label{varepsilon2}
\ell   \left(\mu+\sqrt{\mu^2+p_{\rm i}^2}\right)   J_\ell(p_{\rm i})
= \mu   p_{\rm i}  J_{1+\ell}(p_{\rm i}), \  \ell=0,1,2\dots
\end{equation}
Remember that the solutions $p_{\rm i}(\ell,\mu)$ of these transcendental equations allow us to determine the well depth $v$, which will depend on the parameters $\ell$ and $\mu$, as in the present case $v=\mu-\sqrt{\mu^2+p_{\rm i}^2}$.

\subsubsection{Supercritical states: $\varepsilon= -\mu$}

\begin{figure}[htb]
\centering
\includegraphics[width=0.44\textwidth]{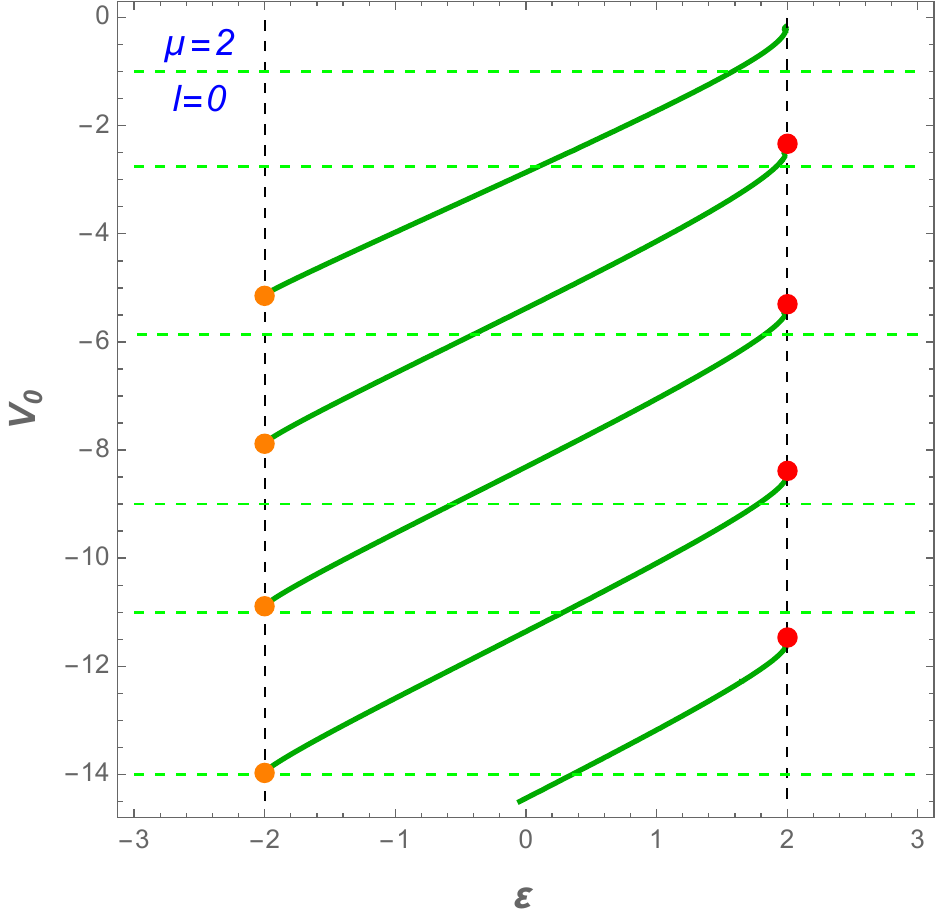}
\qquad
\includegraphics[width=0.44\textwidth]{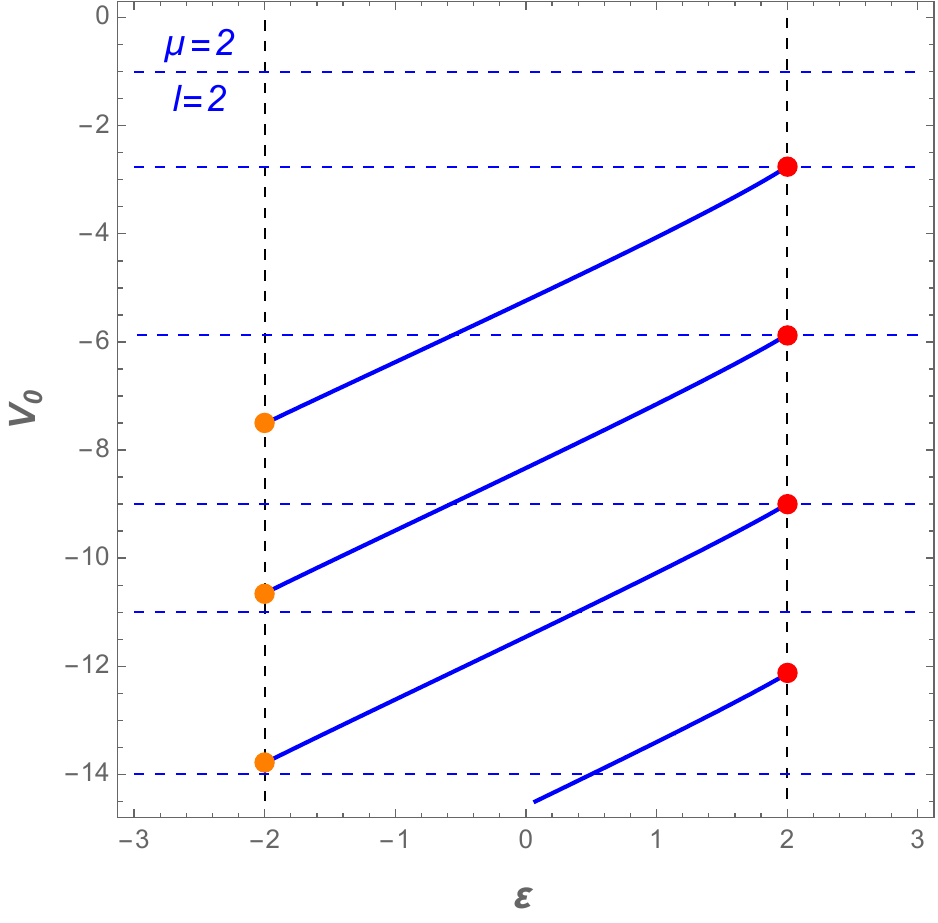}
\caption{\small 
For a particle of mass $\mu=2$: plots of the bound state energy levels obtained from \eqref{secular} when $-2<\varepsilon(v)<2$   for $\ell=0$ (green curves on the left plot) and $\ell=2$ (blue curves on the right plot). They are given as functions of the potential depth in the range $-14<v<0$. The critical states of \eqref{varepsilon2}  are represented by the dots on the right and the supercritical states of \eqref{varepsilon3} by dots on the left sides of the curves. The vertical dashing lines are the limits of
the bound state energies, while the horizontal dashing lines correspond to the values of the well depths $v$ selected for Figure~\ref{fig3}.
\label{fig1}}
\end{figure}

Again, the eigenfunctions in the inner interval $0\leq\rho<1$ are Bessel functions of the first kind \eqref{phileft}, with the following values of the parameters that appear there: $\varepsilon^{+}_{\rm i} = -v>0$ and $p_{\rm i} = \sqrt{v^2+2v\mu}>0$,
while outside, for $\rho>1$, the components satisfy this particular form of equations (\ref{phis})
\begin{equation}\label{phis3}
\left\{
\begin{array}{l}
\displaystyle 
\phi'_{2,{\rm o}} (\rho)+\frac{\ell+1}{\rho}\,\phi_{2,{\rm o}} (\rho) = -2\mu\,\phi_{1,{\rm o}} (\rho) ,
\\ [2ex]
\displaystyle -\phi'_{1,{\rm o}} (\rho) + \frac{\ell}{\rho}\,\phi_{1,{\rm o}} (\rho) = 0 .
\end{array}
\right.
\end{equation}
Then, the solutions bounded in the region $\rho>1$ are  
\begin{equation}\label{phirightsc}
\Phi_{{\rm o}} (\rho,\theta) = a_{\rm o} \left(
\begin{array}{c}
0
\\[1.5ex]
\displaystyle i \rho^{-(\ell+1)} e^{i (\ell+1) \theta}
\end{array}\right) ,
\ \ell= 0,1,2,\dots
\end{equation}
Except the value $\ell=0$, which correspond to a quasi-bound state, the rest, i.e. $\ell\geq 1$, lead to square-integrable wave functions.
The matching condition in the present situation gives rise to the following secular equations:
\begin{eqnarray}\label{varepsilon3}
J_\ell(p_{\rm i}) =0, &&  \ell = 0,1,2, \dots
\end{eqnarray}
which allow to determine the values of $p_{\rm i}(\ell,\mu)$, and therefore those of $v$ for each $\ell$, taking into account that in this case $v=-\mu-\sqrt{\mu^2+p_{\rm i}^2}$.
Some examples of bound energy levels  are given in Figure~\ref{fig1}
along with critical and supercritical
values. In these graphs we can see that for $\ell=0$
bound states appear for any (negative) value of the well depth $v$,
however, for instance, the case $\ell=2$ bound states will only appear for negative values of $v$ lower than  $v=\mu-\sqrt{\mu^2+p_{\rm i,1}^2}$, where $p_{\rm i,1}$ is the first strictly positive root of the transcendental equation \eqref{varepsilon3} ($v=-2.7558$ in the case shown in Figure~\ref{fig1}). 
This fact is due to the centrifugal potential caused by the orbital momentum $\ell$.

Note that from  Figure~\ref{fig1} we can see that for this particular value of mass and for any negative value of $v$ there will
be no more than two bound states for each $\ell$. These two energy levels $\varepsilon_n, \varepsilon_{n+1}$   will always keep in the range $[-\mu,\mu]$ determined by the mass, however the corresponding wave functions $\psi_n^\ell, \psi_{n+1}^\ell$ will have as many zeros as the corresponding to the
excitation number $n$.
Thus, the maximum number of bound states is finite and it depends on the value of $\mu$, but they are the highest excited levels. 

The depth $v$ of the dot modulates the energy of these bound states that belong to the interval 
$[-\mu,\mu]$. 
If the potential depth $v$ is more
negative enough, the lowest bound state may  plunge  into the continuous spectrum of antiparticles, giving rise to the phenomenon of atomic collapse.
This result is quite different from non-relativistic wells, which can have any number of bound states simply by taking more negative depths $v$ without any risk of leaking into the `negative sea'.

Another important remark is the almost linear dependence between the energy $\varepsilon_n$ of each bound level in the interval $[-\mu,\mu]$ and the depth $v$ of the well: $\varepsilon_n(v) \approx \alpha_n v +\beta_n$, where $\alpha_n, \beta_n$ are constants for each  level. 
This linear dependence is more pronounced for higher angular momenta $\ell$, as shown in Figure~\ref{fig1}. In the asymptotic region $|v|>> \mu$, due to the behavior of the Bessel functions, the slope $\alpha_n$ of the levels becomes independent of $n$, while the term $\beta_n$ will be proportional to $n$.

\subsection{Resonances and pure outgoing states}

In a scattering process, a resonance occurs for a real energy $E_r>\mu$  if an incoming wave packet state takes longer time to exit than it should  without the presence of the potential. That is, if it suffers a delay time within the significant range of the potential.  One way to measure these resonances is by computing the phase shift $\delta_\ell$ of incoming and outgoing waves  \cite{dong98,hall04}, as will be done in the next subsection. 
The derivative of this phase shift with respect to energy gives the so-called ``Wigner time delay'' $\tau_\ell$ \cite{wigner,ohanian}, indeed
\begin{equation}\label{time}
\tau_\ell = 2 \, \frac{d\delta_\ell}{d E} .
\end{equation}
The presence of a maximum in this function (specially if it is sharp) is a clear resonance signal.

There is another approach also used to calculate the resonances of a potential well (or barrier) which consists in finding the complex energies  $E=E_R + i E_I$, where the stationary states satisfy {\it purely outgoing boundary conditions}. 
In fact, when these energies are real, in the interval $(-\mu,+\mu)$, they can belong to the discrete spectrum or, in other cases, they can be interpreted as anti--bound states \cite{ohanian,gadella17,gadella16}. 
These characteristics are best appreciated in momentum space, but we will limit ourselves here to the energy picture for simplicity. 
The real part $E_R$ of the resonant energy is then identified 
with the energy of the incident wave, while the imaginary part $E_I$ is related to the resonance delay time as follows: $E_R \approx E_r$, while $\tau_\ell\propto 1/E_I$. 
The consistency  of these two criteria (Wigner time delay and complex energies) to detect resonances will be checked in the next two subsections.

\subsubsection{Scattering states, phase shifts, and resonances \label{scatt}}

\begin{figure}[h!]
\centering
\includegraphics[width=0.44\textwidth]{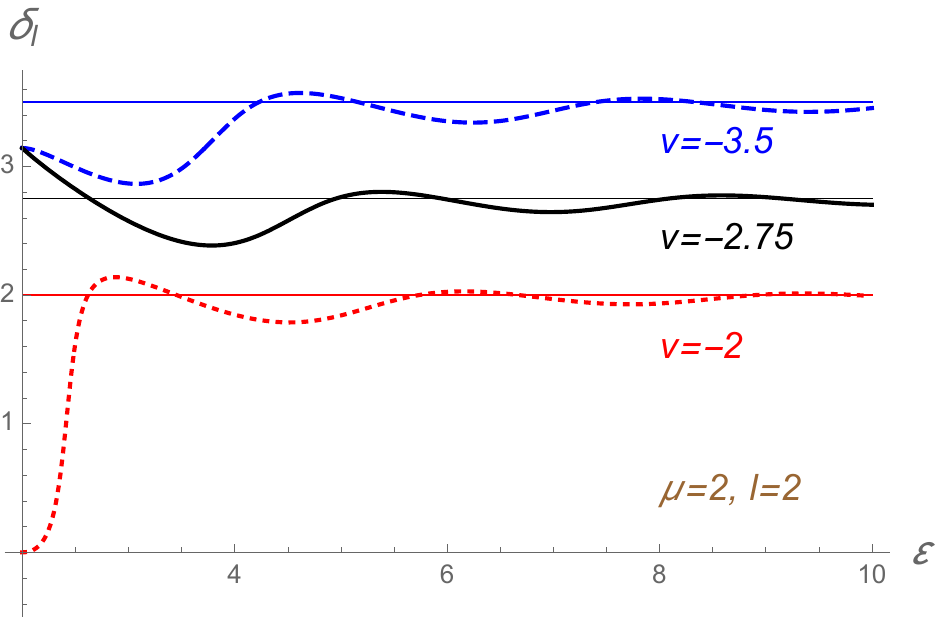}
\qquad\qquad
\includegraphics[width=0.44\textwidth]{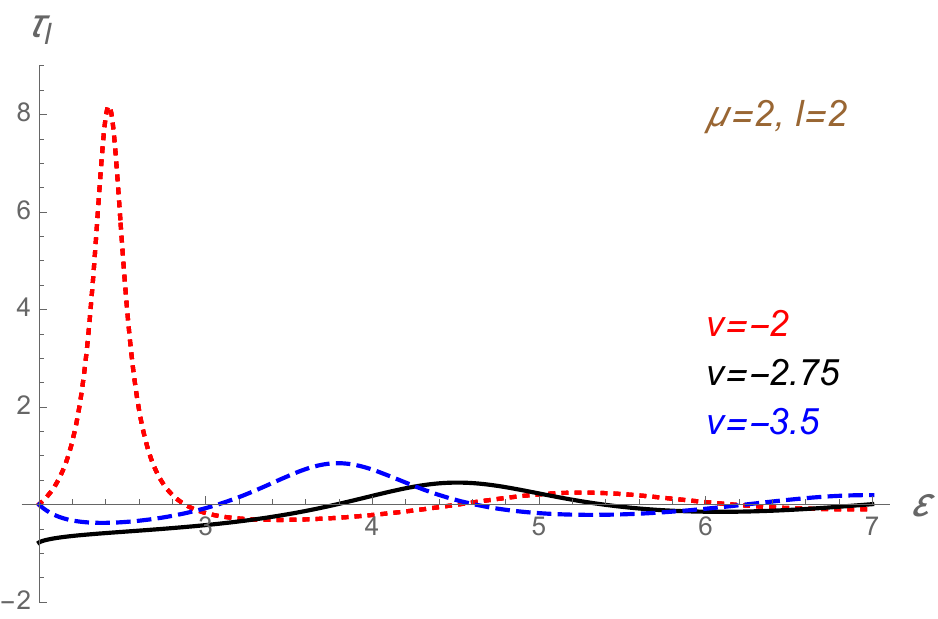}
\caption{\small On the left, a graph of the phase shift $\delta_\ell(\varepsilon)$ for the scattering of a particle with $\mu=2$, $\ell=2$, and three  values of the potential depth: (a) $v=-3.5$, just after trapping a bound state  (blue dashed line), (b) $v = -2.75$,  the trapping value  (black solid line), and (c) $v=-2$, before trapping a bound state  (red dotted line). Note that due to the capture of a bound state taken from the continuum, there is a phase jump (see the phases of $v=-2.75$ and $v= -3.5$ with respect to that of $v=-2$). 
The horizontal lines are the limits of the phases when $\varepsilon \to
\infty$ from (\ref{lim}). Remark that hereafter  the critical or supercritical values of the potential will be given with two decimals in order to simplify the notation (as it is the case of $v = -2.75$).
On the right, a plot of the corresponding Wigner time delays for the same values of $v$. 
The maxima of the curves show the existence of resonances for particular values of the scattering energy.
\label{fig2}}
\end{figure}

In order to study the scattering states and their phase shifts, 
it is better to use in the outer region the basis
$\{J_\ell,Y_\ell\}$. The asymptotic behavior
of these functions for large values of $\rho$ is the following \cite{abramowitz}:
\begin{equation*}
J_\ell(p_{\rm o}\, \rho) \sim \sqrt{\frac{2}{\pi p_{\rm o}\, \rho }}
\cos(p_{\rm o}\, \rho-\ell \pi/2 - \pi/4) ,
\qquad
Y_\ell(p_{\rm o}\, \rho) \sim \sqrt{\frac{2}{\pi p_{\rm o}\, \rho }}
\sin(p_{\rm o}\, \rho-\ell \pi/2 - \pi/4) .
\end{equation*}
Then, the spinors of the scattering states take the form \eqref{phileft}
in the inner region ($0\leq\rho<1$), and the following one in the outer region ($\rho>1$)
\begin{equation*}
\Phi_{\rm o}(\rho,\theta) =\left(
\begin{array}{c}
\bigl[ A\,  J_\ell(p_{\rm o}\,\rho) + B\,  Y_\ell(p_{\rm o}\,\rho) \bigr] \, e^{i \ell \theta}
\\[1.5ex]
\displaystyle i \frac{p_{\rm o}}{\varepsilon_{\rm o}^+} 
\bigl[ A\,   J_{\ell+1}(p_{\rm o} \rho) + B\,   Y_{\ell+1}(p_{\rm o} \rho) \bigr] \,  e^{i (\ell+1) \theta}
\end{array}\right).
\end{equation*}
If for convenience we choose the form of the arbitrary constants $A$ and $B$ as $A = a \cos \delta_\ell$,  $B = -a \sin \delta_\ell$,
the asymptotic behavior of the spinor when $\rho\to \infty$ is
\begin{equation*}
\Phi_{\rm o}(\rho,\theta) \sim
\sqrt{\frac{2}{\pi p_{\rm o}  \rho }}
\left(\begin{array}{cc}
\cos(p_{\rm o}  \rho-\ell \pi/2 - \pi/4+\delta_\ell) \, e^{i \ell \theta}
\\[1.5ex]
i \frac{p_{\rm o}}{\varepsilon_{\rm o}^+} \sin(p_{\rm o}  \rho-\ell \pi/2 - \pi/4+\delta_\ell)\,
  e^{i (\ell+1) \theta}
\end{array}\right) .
\end{equation*}
The quantity $\delta_\ell$ is the phase shift due to the presence of the potential near the origin of coordinates, and it appears in the outer spinor wave function:
\begin{equation*}
\Phi_{\rm o}(\rho,\theta)=A \left(
\begin{array}{c}
\bigl[ J_\ell(p_{\rm o} \rho) -\tan \delta_\ell  Y_\ell(p_{\rm o} \rho) \bigr]   e^{i \ell \theta}
\\[1.5ex]
\displaystyle i \frac{p_{\rm o}}{\varepsilon_{\rm o}^+} 
\bigl[ J_{\ell+1}(p_{\rm o} \rho) -\tan \delta_\ell  Y_{\ell+1}(p_{\rm o} \rho) \bigr]   e^{i (\ell+1) \theta}
\end{array}\right) .
\end{equation*}
The continuity condition of the spinor $\Phi(\rho,\theta)$ at $\rho=1$ leads to
\begin{equation*}
\frac{\varepsilon_{\rm i}^+\, J_\ell(p_{\rm i})}{p_{\rm i}\, J_{\ell+1}(p_{\rm i})}
=
\frac{\varepsilon_{\rm o}^+( J_\ell(p_{\rm o}) -\tan \delta_\ell\, Y_\ell(p_{\rm o}))}{
p_{\rm o} (J_{\ell+1}(p_{\rm o}) -\tan \delta_\ell\, Y_{\ell+1}(p_{\rm o}))},
\end{equation*}
from where we obtain the explicit value of the phase $\delta_\ell$:
\begin{equation}\label{delta}
\tan \delta_\ell(\varepsilon) = 
\frac{\varepsilon_{\rm i}^+ \, p_{\rm o} \, J_\ell(p_{\rm i})J_{\ell+1}(p_{\rm o})
-\varepsilon_{\rm o}^+ \, p_{\rm i}\, J_{\ell+1}(p_{\rm i})J_{\ell}(p_{\rm o})}
{\varepsilon_{\rm i}^+ \, p_{\rm o} \,  J_\ell(p_{\rm i})Y_{\ell+1}(p_{\rm o})
-\varepsilon_{\rm o}^+ \, p_{\rm i}\, J_{\ell+1}(p_{\rm i})Y_{\ell}(p_{\rm o})} .
\end{equation}
Once  the potential depth $v$ is set, the phase shift $\delta_\ell$ will depend on the energy $\varepsilon$, taking into account \eqref{varepsilon} and \eqref{ps}. 
Also,  for a potential well with depth $v$ it is easy to show that at the high energy limit (note that the constraint \eqref{conditionbound} is no longer valid now)  the phase shift is
\begin{equation}\label{lim}
\lim_{\varepsilon\to\infty}\tan \delta_\ell(\varepsilon) = -\tan v .
\end{equation}
The values for which the derivative with respect to the energy of \eqref{delta} is maximum correspond to resonances, since this will mean that with this energy the time that a wave packet spends inside the well  \eqref{time} will be greater than it should. 
It also turns out that as we increase the depth of the well we trap bound states that leave the continuum. 
In this process, when a value of $v$ is reached so that a new bound state is captured, the corresponding phase shift  undergoes an abrupt change, increasing by $\pi$ (which is the content of Levinson's theorem adapted to the relativistic plane \cite{hall04,dong98}). These features
are shown in Figure~\ref{fig2}.

\subsubsection{Complex resonances and outgoing states \label{complex}}

In the scattering process discussed in the previous subsection, we had an incoming wave and an outgoing wave, and we calculated the phase shift of these waves due to the potential near the origin.
Next, we will look for energy values $\varepsilon$ such that we have a pure outgoing wave.
This situation may not have a physical realization, but it will provide us with useful information. 
In fact, this condition in general will be satisfied for
complex energies: $\varepsilon = \varepsilon_R + i \varepsilon_I$. 
Therefore, states with pure outgoing boundary conditions must satisfy the following conditions: 
\begin{enumerate}
\item[(i)] 
when $\rho\to0$
the spinor $\Phi_{\rm i}(\rho,\theta)$ must be kept bounded,
\item[(ii)] 
when $\rho\to \infty$ each component must behave
as the first Hankel function (\ref{asympH1}), and the spinor $\Phi_{\rm o}(\rho,\theta)$ as
in (\ref{phiright}). 
\end{enumerate}
In other words, the purely outgoing wave conditions are the same as the bound state
conditions \eqref{secular}, except that now what we want to find are the complex solutions  $\varepsilon = \varepsilon_R + i \varepsilon_I$ of this secular equation, and therefore the corresponding eigenfunctions may diverge when $\rho\to \infty$.

\begin{figure}[tb]
\centering
\includegraphics[width=0.44\textwidth]{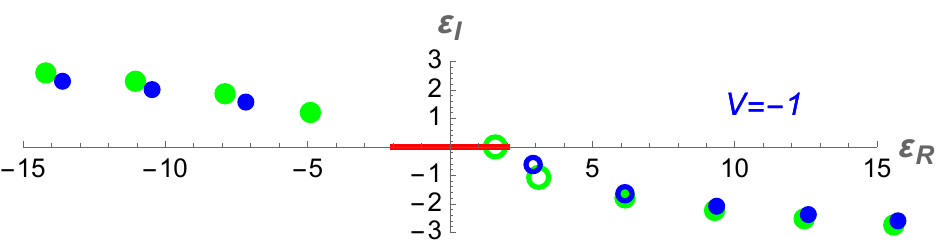}
\qquad \qquad
\includegraphics[width=0.44\textwidth]{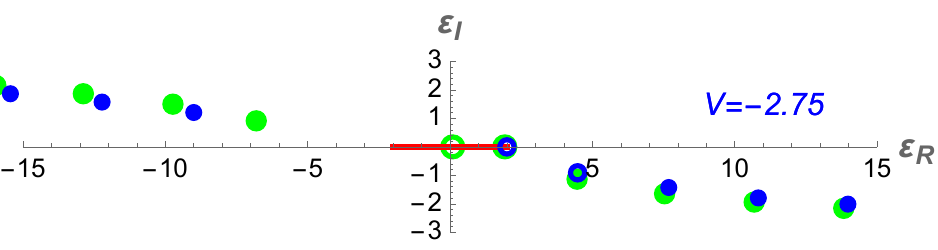}
\\[1.5ex]
\includegraphics[width=0.44\textwidth]{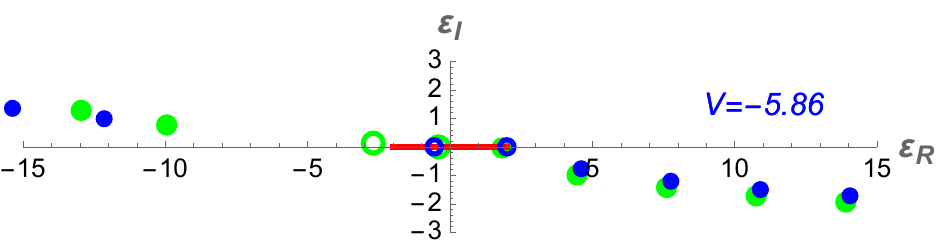}
\qquad\qquad
\includegraphics[width=0.44\textwidth]{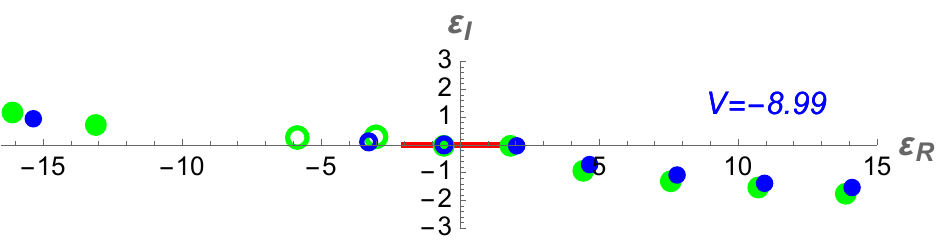}
\\[1.5ex]
\includegraphics[width=0.44\textwidth]{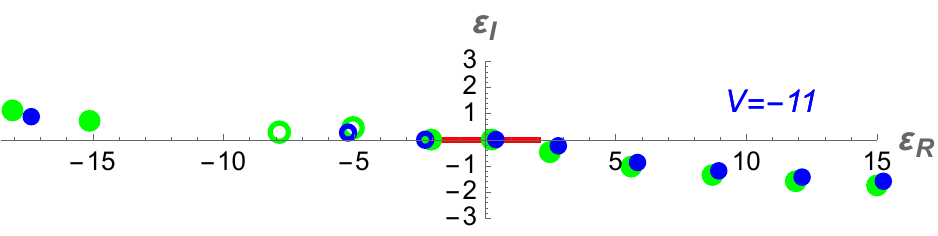}
\qquad  \qquad
\includegraphics[width=0.44\textwidth]{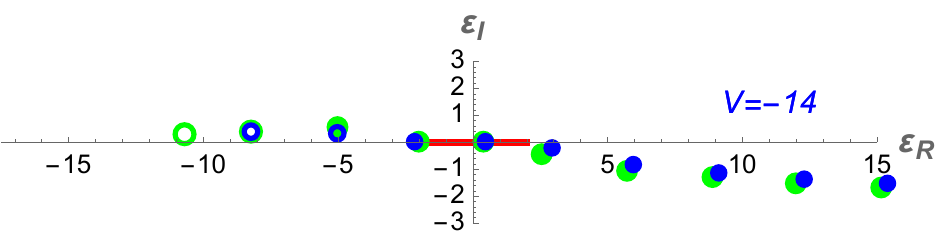}
\caption{
\small Resonances in the complex plane $\varepsilon=\varepsilon_R+i \varepsilon_I$ for particles with mass $\mu=2$, angular momenta $\ell=2$ (blue dots) and $\ell=0$ (green dots), are represented 
as functions of  the  well depth $v$.
There are six plots for six values of $v$ (indicated in each graphic). 
The thick red interval $(-2,2)$ on the real axis represents the possible bound energies.
The evolution of the resonances as the depth of the potential well goes from the initial value $ v=-1 $ up to the final one $v=-14$ is clearly observed by the change of position of the resonances drawn as circles (green for $\ell=0$, blue for $\ell=2$).
At $v= -2.75, -5.86, -8.99$ the potential well captures a new bound state, with $\ell =2$, of energy $\varepsilon=2$ (recall that we use two decimals for numerical values of critical potentials).
In general, as $v$ becomes more negative, the set of resonances moves from energies with positive values of $\varepsilon_R$ towards energies with negative values $\varepsilon_R$. The complex resonance energies
for $\ell=2$ are slightly closer to the real axis (therefore stronger) than those for $\ell=0$.
\label{fig3}}
\end{figure}

Obviously, the resonances depend on the depth $v$, in the same way that the energies of the bound states also depend on $v$, as shown in Figure~\ref{fig1}. As the results must necessarily be obtained numerically or graphically from \eqref{secular}, we will choose, as an example, a particle with mass $\mu=2$ and follow its trajectory $\varepsilon(v) = \varepsilon_R(v) + i \varepsilon_I(v)$ as a function of $v$. 
We start from a resonance such that $\varepsilon_R(v_0)> \mu =2$,  $\varepsilon_I(v_0)\neq 0$, and little by little we decrease the depth of the potential well until we reach the value  $v_1$ for which we have precisely the first bound state: $\varepsilon(v_1) = \varepsilon_R(v_1) = \mu$, $\varepsilon_I(v_1)= 0$.
As we continue to decrease the value of $v$, the bound energy 
decreases to the minimum bound eigenvalue $\varepsilon(v_2) = \varepsilon_R(v_2) = -\mu=-2$. 
Below this value $v_2$, the bound energy will again change into a complex
resonance $\varepsilon(v_3) = \varepsilon_R(v_3) + i \varepsilon_I(v_3)$, with $\varepsilon_R(v_3)< -2$ and $\varepsilon_I(v_3)\neq 0$. 
The whole process can be followed in Figure~\ref{fig3}. We observe that, as shown in that figure, the resonances for the non-zero momentum, $\ell=2$, are higher but of the same order as those with zero momentum $\ell=0$. This is shown in greater detail in Figure~\ref{fig3}, as explained below.

\begin{figure}[h!]
\centering
\includegraphics[width=0.44\textwidth]{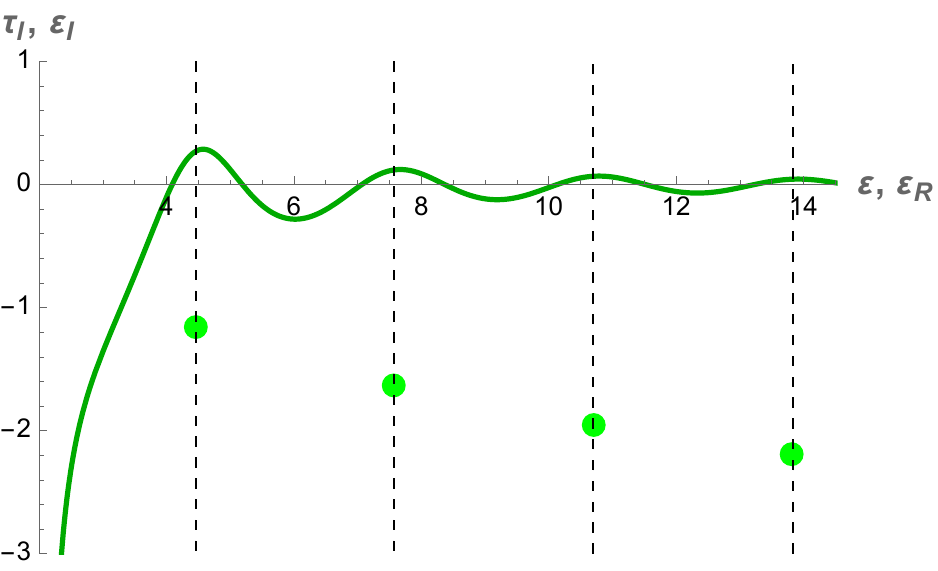}
\qquad\qquad 
\includegraphics[width=0.44\textwidth]{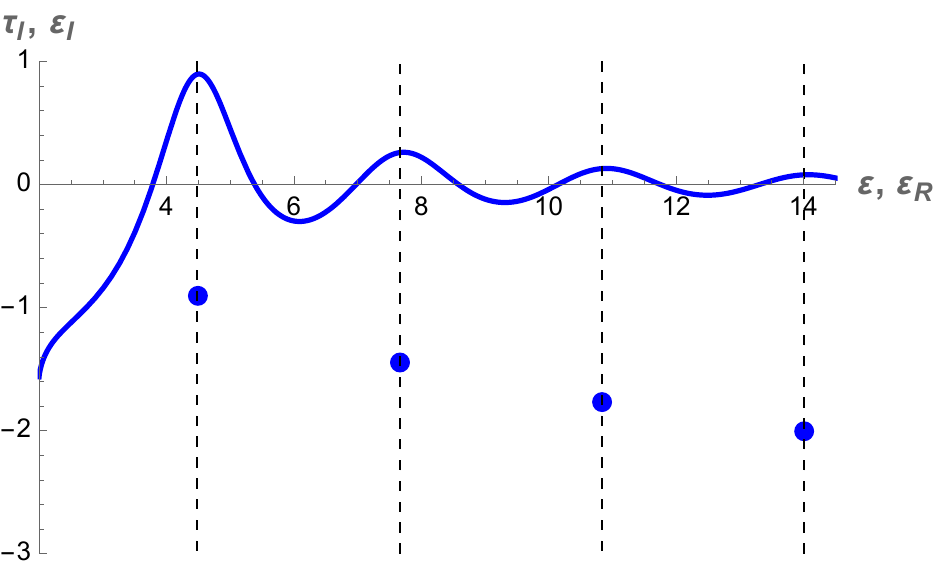}
\caption{
\small The left graphic in green is for $m=2,\ell=0$, the right in blue for $m=2,\ell=2$, in both plots the potential depth is $v=-2.75$ (see Figure \ref{fig3}). The dots 
represent complex resonances with coordinates 
$(\varepsilon_R,\varepsilon_I)$. The continuous curves are for the Wigner time delay as a function of the real part of the energy. 
The maxima of the Wigner time delay take place at energies which have a good agreement with the real part of the resonance energies $\varepsilon_R$
as shown in both graphics by the dashing vertical lines. However, in the left one (for $\ell=0$) the coincidence is not so good  (this may be due to the fact that for $\ell=0$ the critical `bound state' is not square integrable).
The resonances for $\ell=2$ are stronger  than for $\ell=0$, but both have the same order of magnitude.   
\label{fig4}}
\end{figure}

The relation between the Wigner time delay $\tau_\ell$ and the complex energies is shown in Figure~\ref{fig4}: the dots correspond to resonances in the energy representation and the curves represent the Wigner time delay as a function of energy. For $\ell>0$, we see that the real part of the resonances perfectly coincides with the peak of the time delay, but for $\ell=0$ this correspondence is not so good due to the quasi-bound character of the  critical state if $\ell=0$, as already mentioned in (\ref{phirightc}).

\section{Massless two-dimensional Dirac particles: Bound states and resonances}

In this section we will analyze the bound states and resonances of the problem under study when the particles are assumed to be massless, something that can be seen as a limit of the treatment given in the previous section for massive particles. However, we will see that the behavior in massive and massless cases has important differences that we will highlight below.

\subsection{Bound states}

The potential well has the same shape as in the
case of non-zero masses (\ref{potentialwell}). The equations for the components $\phi_1,\phi_2$ also have the same form as their massive analogs  (\ref{phis}), although now the mass disappears from the equations as $\mu=0$. 
Consequently, the energy constants given in (\ref{varepsilon}) and (\ref{ps}) become
\begin{equation}\label{varepsilonB}
\varepsilon_{\rm i}^\pm=\varepsilon -v ,\quad 0\leq\rho<1\,,
\qquad
\varepsilon_{\rm o}^\pm=\varepsilon ,\quad \rho>1,
\end{equation}
and
\begin{equation}\label{psB}
p_{\rm i}= \pm (\varepsilon-v) ,\quad 0\leq\rho<1\,,
\qquad 
p_{\rm o}=  \pm \varepsilon ,\quad \rho>1,
\end{equation}
where the appropriate sign in (\ref{psB}) must be chosen in each of the two cases (interior and exterior) in order to obtain correct boundary conditions for the solutions.

By the same arguments about the asymptotic behavior of the wave functions of the massive case \eqref{conditionbound},  in the current situation the bound states must have zero energy $\varepsilon=0$, that is, they coincide with the critical and supercritical states. 
To find them we concentrate on the outer region $\rho>1$, because in the inner region the solutions are the same as in the case $\mu\neq 0$. In the outer region, since $\varepsilon_{\rm o}^\pm=0$, the equations (\ref{phis}) become
\begin{equation}\label{phis2}
\phi_2' +\frac{\ell+1}{\rho}\,\phi_2 = 0 \,,
\qquad  
-\phi_1' + \frac{\ell}{\rho}\,\phi_1 = 0\, .
\end{equation}
Notice that in these two equations there is the symmetry $\ell \to -(\ell +1)$ and changing
the components $\phi_1 \to \phi_2$, due to the fact that $\mu=0$.
The solutions to these equations are (see also \cite{portnoi19})
\begin{equation*}
\Phi_{\rm o}(\rho,\theta)  = \left(\begin{array}{c}
c_1\, \rho^\ell  \, e^{i \ell \theta}
\\[1.ex]
i\,c_2\,\rho^{-(\ell+1)} \, e^{i (\ell+1) \theta}
\end{array}\right),
\  \ell=0,\pm1,\pm2,\dots
\end{equation*}
where $c_1,c_2$ are arbitrary integration constants.  
Thus, if either $\ell>0$ or $\ell<-1$, the physically acceptable bound states are described by 
\begin{eqnarray*}
\Phi_{\rm o}(\rho,\theta)  = \left(\begin{array}{c}
0
\\[1.ex]
i\,c_2\,\rho^{-(\ell+1)} \, e^{i (\ell+1) \theta}
\end{array}\right),\quad \ell>0,
\qquad 
\Phi_{\rm o}(\rho,\theta) = \left(\begin{array}{c}
c_1\, \rho^\ell  \, e^{i \ell \theta}
\\[1.ex]
0
\end{array}\right),\quad \ell<-1,
\end{eqnarray*}
which go to zero as $\rho\to \infty$ and are square integrable. For
the cases $\ell=0,-1$ the corresponding solutions, although also vanish at infinity,
are not square integrable, and thus they do not correspond to `true' bound states.

\begin{figure}[h!]
\centering
\includegraphics[width=0.13\textwidth]{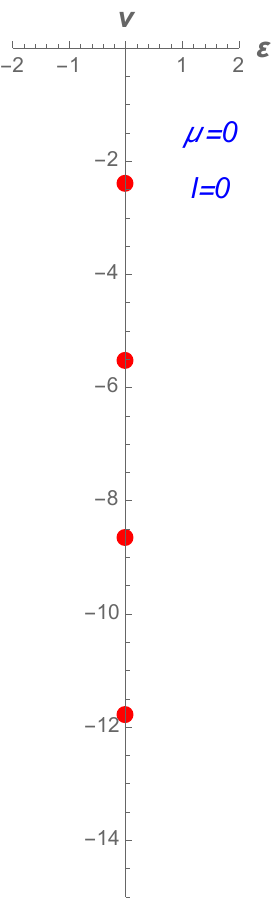} 
\qquad \qquad\qquad
\includegraphics[width=0.13\textwidth]{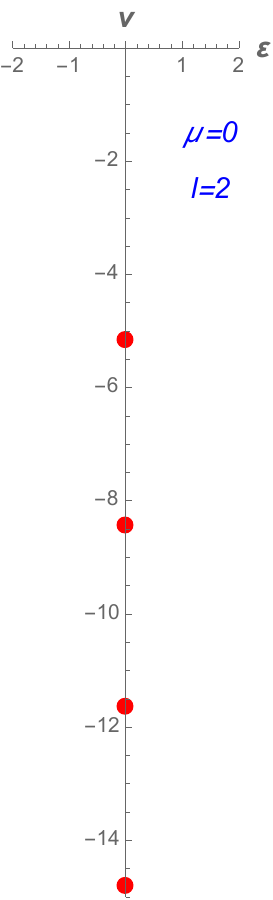} 
\caption{
\small The red dots are the well depth $v$ values for which there are bound states  with energy $\varepsilon =0$ for massless  particles ($\mu=0$) from \eqref{varepsilon4}: on the left the case with $\ell=0$, on the right the case with $\ell=2$. These two graphs correspond to the limit $\mu\to0$ of the massive cases, like the ones in Figure~\ref{fig1} for $\mu=2$.
\label{fig5}}
\end{figure}

The formula for the potential depth $v$ is the same as
\eqref{varepsilon3} taking the limit $\mu\to 0$,
\begin{eqnarray}\label{varepsilon4}
J_\ell(p_{\rm i}) =0, \  \ell \geq  0,
\quad
 J_{1+\ell}(p_{\rm i})= 0, \ \ell<0,
 \quad p_{\rm i}= |v|.
\end{eqnarray}
Then, the roots of $J_\ell(p_{\rm i})=0$, give us the values for $v$ corresponding the bound states. In Figure~\ref{fig5} two graphics represent the first of these values corresponding to the angular momenta $\ell=0$ and  $\ell=2$.

It is worth to mention a useful method to investigate the spectrum of some systems reduced to one dimension in graphene
called variable phase \cite{miserev}. It supplies a new insight on some problems not clearly explained with standard methods.

\subsection{Scattering states and resonances}

\begin{figure}[htb]
\centering
\includegraphics[width=0.44\textwidth]{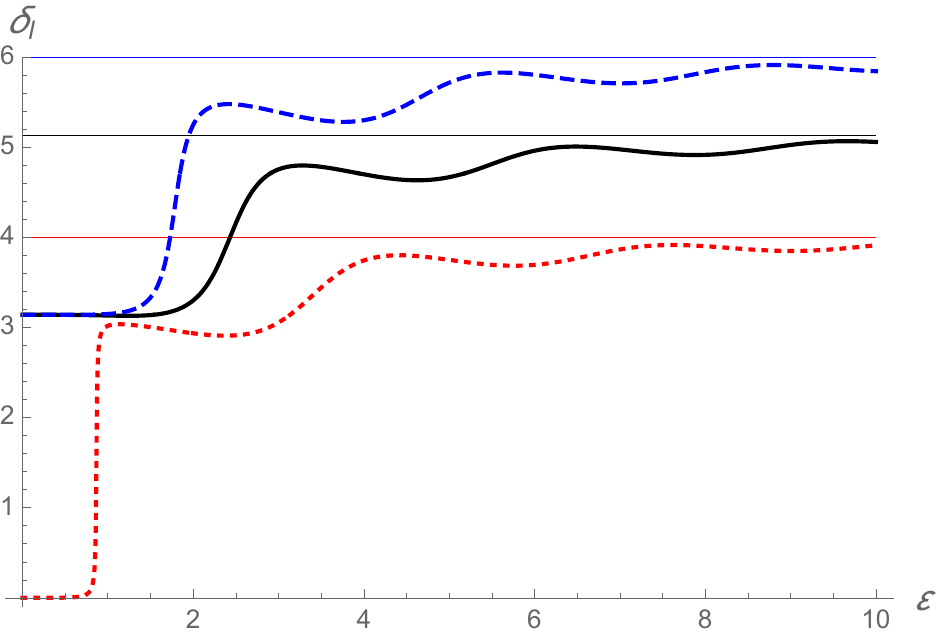}
\qquad\qquad
\includegraphics[width=0.44\textwidth]{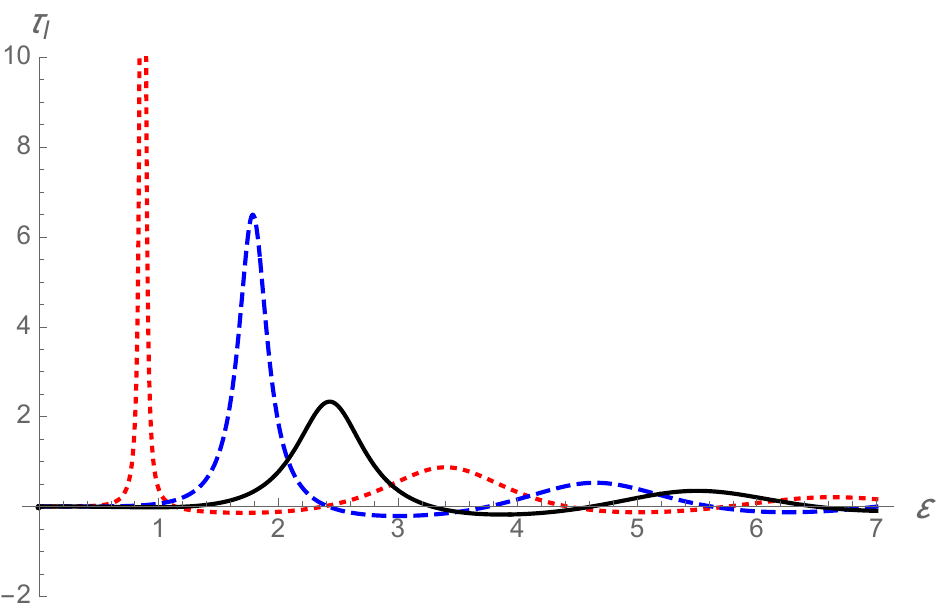}
\caption{
\small For massless particles and $\ell=2$, the graph on the left  represents the phase shift depending on the incident energy for three values of  potential depth  $v=-4$ (dotted), $v=-5.13$ (solid), and $v=-6$ (dashed); in the graph to the right, the corresponding Wigner time delay is plotted as a function of energy for the same three potential depth values. The highest resonances (the red dashed in the graphics, $v=-4$) are obtained for depths slightly less than a critical value, in this case $v=-5.13$. 
\label{fig6}}
\end{figure}

With respect to scattering states with positive energy $\varepsilon>0$, the situation is completely similar to Subsection~\ref{scatt} for massive particles. 
The phase shift of the scattering states is calculated using (\ref{delta}), whose limit value when $\varepsilon\to \infty$ is (\ref{lim}). In Figure~\ref{fig6}, some examples of phase shift of scattering states and their derivatives with respect to energy (interpreted as Wigner time delays) are shown.
The resonances are the  $\varepsilon_r(v)$ values for which the Wigner time delay reaches a maximum. 
We have verified numerically that even for $\ell=0$ there are time delays and resonances. However, for $\ell>0$ they become quite strong, especially for energies close to capturing a bound state.

On the left of Figure~\ref{fig6} we have represented  the phase shifts for a massless charge with angular momentum $\ell=2$ for three values of $v$: $-4$, $-5.13$, and $-6$. The special value $v = -5.13$ corresponds to the capture of a bound state with $\varepsilon=0$.
For the value $v=-4$ (which is a bit above  $v=-5.13$) the phase shifts undergo a strong change at a certain value $\varepsilon$ near
$\varepsilon=0$ (dotted curve). 
However, for the capture value $v=-5.13$ (solid curve) or for a slightly lower value ($v=-6$, dashed curve), the phase shifts are smoother and start with a jump of $\pi$ in $\varepsilon=0$ (according to  Levinson's theorem for massless particles \cite{dong98}). 
Furthermore, in the limit $\varepsilon\to \infty$, it is seen that
these phase shifts tend to the corresponding value $|v|$.
To the right of Figure~\ref{fig6} we have represented  the derivatives of these three phase shifts, which are identified with Wigner time delays. 
The potential $v=-4$ (a little above the capture
value $v=-5.13$) has a very high maximum of the time delay $\tau$, reached at a certain value $\varepsilon_r$ that we identify with a strong resonance.
However, for the capture potential $v=-5.13$ or slightly lower values ($v=-6$), the maxima of time delays are much lower, corresponding to weak resonances. In summary, the behavior of scattering states is reasonable according to the non-relativistic nonzero mass theory on phase shift and Wigner time delay.

\begin{figure}[htb]
\centering
\includegraphics[width=0.44\textwidth]{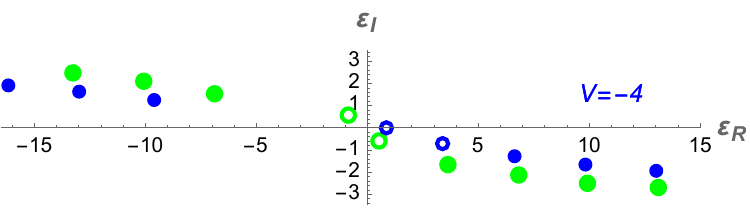}
\qquad \qquad
\includegraphics[width=0.44\textwidth]{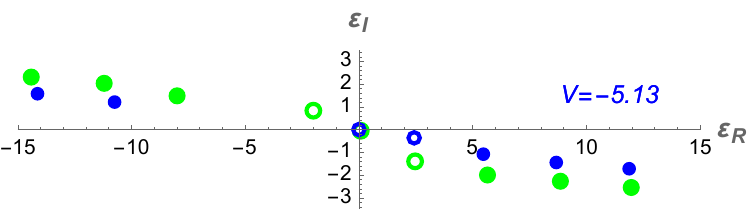}
\\[1.5ex]
\includegraphics[width=0.44\textwidth]{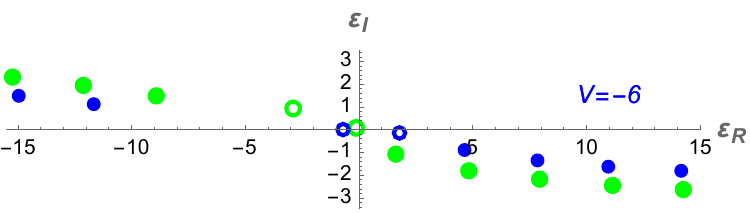}
\qquad\qquad 
\includegraphics[width=0.44\textwidth]{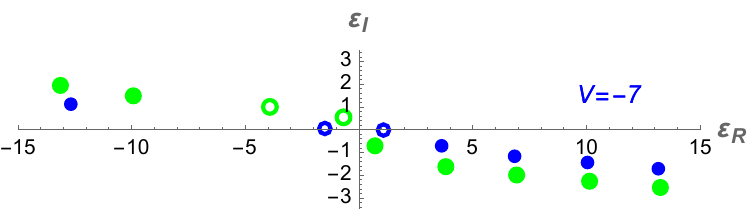}
\\[1.5ex]
\includegraphics[width=0.44\textwidth]{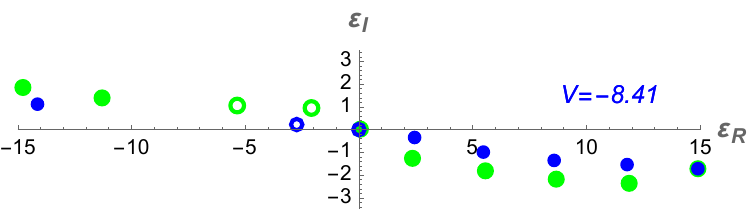}
\qquad \qquad
\includegraphics[width=0.44\textwidth]{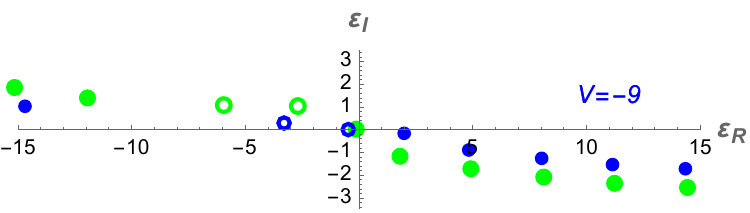}
\caption{\small Resonances for a massless particle with angular momenta $\ell=0$ (green dots) and $\ell = 2$ (blue dots). For higher value $\ell=2$ they are much closer to the real axis than the ones with null
angular momentum $\ell=0$. As a consequence, the corresponding blue resonances are much larger (an order of magnitude) than the green ones. This is a differential behavior with massive Dirac particles.
\label{fig7}}
\end{figure}

As we have already mentioned, another way to define (complex) resonances is through  complex
eigenvalues of energy corresponding to eigenfunctions that satisfy purely outgoing boundary conditions. 
Then, we must look for 
complex solutions $\varepsilon(v)=\varepsilon_R+i\,\varepsilon_I$ of equation (\ref{secular}), but now having in mind that the mass vanishes and therefore  also the simplifications (\ref{varepsilonB})-(\ref{psB}) apply. According to (\ref{asympH1}) the asymptotic behaviour of the Hankel function, which concern us, is
\begin{equation}\label{asympH1b}
H_\ell^{(1)}(p_{\rm o} \rho) \sim  
e^{i(p_{\rm o}  \rho -\ell\pi/2-\pi/4)} .
\end{equation}
In our case, from (\ref{psB}), $p_{\rm o}=  \pm \varepsilon= \pm \varepsilon_R \pm i \varepsilon_I$.
The wave will be outgoing as far as $\pm \varepsilon_R >0$. This condition is fulfilled if
we choose the positive sign for $\varepsilon_R >0$ (for the resonances in the right hand complex plane);
and the negative sign for $\varepsilon_R <0$ (for the resonances in the left hand complex plane).
In conclusion, we obtain the following secular equation, depending on the sign of the real part of the complex energy $\varepsilon$, 
\begin{equation}\label{secularbis}
    J_\ell(\varepsilon_R+i\,\varepsilon_I -v) \, H^{(1)}_{\ell+1} \Bigl( |\varepsilon_R|+i\,\varepsilon_I\, \text{sign}(\varepsilon_R)\Bigr) - \text{sign}(\varepsilon_R)\, J_{\ell+1}(\varepsilon_R+i\,\varepsilon_I -v) \, 
  H^{(1)}_\ell \Bigl( |\varepsilon_R|+i\,\varepsilon_I\, \text{sign}(\varepsilon_R)\Bigr)   =0 .
\end{equation}
Some solutions of this equation are shown  in  Figure~\ref{fig7} for six values of $v$.

As we have already shown in the previous subsection, the bound states, with $\varepsilon=0$, appear only for some special values of the well depth and are critical states. In Figure~\ref{fig7}, such bound states occur for $v=-5.13$ and for $v=-8.65$ (following our convention of two decimals). The resonances shown in Figure~\ref{fig6} correspond to the well depth values  $v=-4,-5.13,-6$, which are part of Figure~\ref{fig7}.
For the value $v=-4$, the resonance is closest to the origin, as shown in Figure~\ref{fig7} with a blue circle, and is represented by the maximum of the dotted curve in Figure~\ref{fig6} (right); this is a strong resonance. 
For the next value $v=-5.13$, that resonance becomes a bound state with zero energy and the next  resonance is represented at $\varepsilon_R\sim 2.50$ by the maximum of the solid black curve in Figure~\ref{fig6}; in this case the resonance is much weaker.
Finally, for the value $v=-6$, the first resonance is closer to the origin, $\varepsilon \sim 1.90$, and becomes slightly stronger than the previous one, as seen by the maximum of the dashed curve in Figure~\ref{fig6}.

\begin{figure}[htb]
\centering
\includegraphics[width=0.44\textwidth]{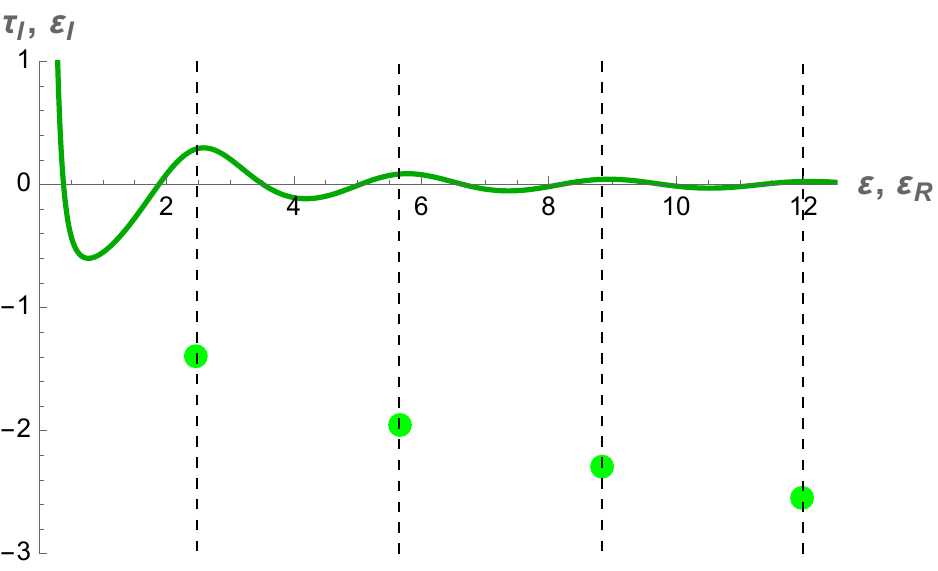}
\qquad\qquad  
\includegraphics[width=0.44\textwidth]{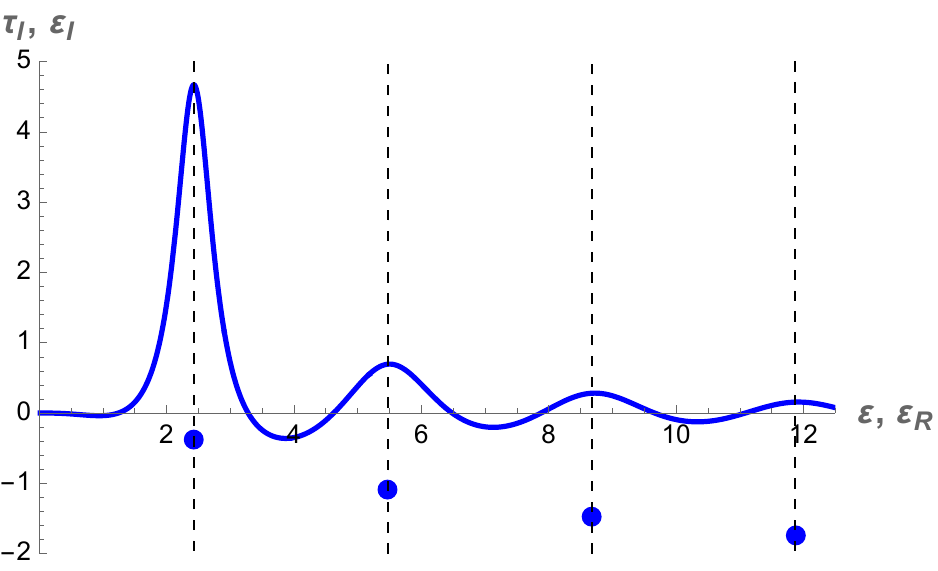}
\caption{
\small On the left, in green, plot of Wigner time delay as a function of energy (solid curve) and complex resonances $\varepsilon= \varepsilon_R+ i \varepsilon_I$ (represented by dots at points $(\varepsilon_R,\varepsilon_I)$)  for massless particles with $\ell=0$. The shape of time delay fails close to $\varepsilon \sim 0$ due to the bad behaviour of the bound state at $\varepsilon=0$.
On the right, in blue, the same plots for $m=0$ and $\ell=2$. The depth of the well  for both cases is $v=-5.13$ (capture value of a bound state for $\ell=2$, see Figure~\ref{fig7}).
\label{fig8}}
\end{figure}

If both approaches to resonance phenomena correspond to the same physical
concept, described by different properties, then we should have
$\varepsilon_r(v)\approx \varepsilon_R(v)$, where $\varepsilon_r(v)$ is the real energy of the scattering state and $\varepsilon_R(v)$ is the real part of a complex resonance. 
The imaginary part $\varepsilon_I$ is inversely proportional to the time delay. From Figure~{\ref{fig8}}, we can  see  the close relationship between the complex resonances of Figure~\ref{fig7} (represented by green dots for $\ell = 0$ and blue dots for $\ell=2$) with coordinates $(\varepsilon_R,\varepsilon_I)$ and the Wigner time delay (represented by a green curve for $\ell=0$, or a blue one for $\ell=2$). 
We observe that the first coordinate $\varepsilon_R$  of the dots
(representing complex resonances) is very close to the values of the maxima $\varepsilon_r$ of the Wigner time delays (this is shown by the dashed vertical lines), specially for the value  $v=-4$ which is close (from above) to the value $v=-5.13$ corresponding to the capture of a bound state, as shown in Figure~\ref{fig7}. If we compare Figure~\ref{fig8} and Figure~\ref{fig4}, we can see that angular momenta affect much more
to massless resonances than to massive ones.

We must point out the differences in the evolution of the resonances of the  massive particles in Figure~\ref{fig3} and those of the massless case in Figure~\ref{fig7}. It is clear that for deeper wells, both the resonances
and the bound states move towards deeper values of (the real part of) the energy. 
This is true for both massive and massless Dirac particles. However, angular momentum has little influence on massive particles but, on the contrary, it is quite important for massless particles. In fact, for $\ell=0$, the resonances are very weak while for $\ell=2$ they are quite strong for $m=0$, as we saw in detail in Figures \ref{fig4}-\ref{fig8}.

\section{Conclusions}

The aim of this article was the search of specific confining properties of both massive and massless Dirac particles in the so-called Dirac materials, including graphene, under the influence of cylindrical electric quantum dots. We have paid attention to the role of angular momentum and mainly to the resonance states in the confinement process.   Next, we will briefly comment on the main novelties as well as the differences found with respect to previous treatments of similar problems.

The spectrum of bound states of a massive Dirac fermion, $\mu\neq 0$, with zero and non-zero angular momentum, is plotted on appropriate global plots in Figure~\ref{fig1}, where the energy levels are represented in terms of the depth  $v$ of the potential well on the vertical axis and the energy values on the horizontal axis.
The most outstanding characteristics shown in the graphs of Figure~\ref{fig1} are the following:

\begin{itemize}
\item[(i)]  The range of bound energies for any value $v$ of the potential is $-\mu\leq \varepsilon\leq \mu$. Then, we have represented the capture of bound levels through the critical points  and the collapse of levels at the supercritical points of $v$. These two key points were determined and their associated eigenfunctions were calculated. 

\item[(ii)]  Due to the phenomenon of atomic collapse, the number of discrete energy levels as a function of  potential height does not increase monotonically (as in the non relativistic case) but this number remains bounded (new levels captured from the positive continuous spectrum are compensated for the lost levels  in the negative continuous spectrum through collapses). This is verified by  counting the number of cuts of horizontal dashing lines at different values of $v$  with the discrete energy level curves in Figure~\ref{fig1}. 

\item[(iii)]  Angular momentum affects these discrete levels: the higher angular momentum, the deeper wells are needed to capture the bound states of the continuum, which is reasonable since the effective centrifugal terms that increase with $\ell$.

\item[(iv)]  Our plots of  discrete energy curves plus critical points, which are quite useful in understanding the global behavior of bound states, are not provided in most of the previous references dealing with this problem, nor in articles discussing confinement in a Coulomb impurity \cite{pereira07,shytov07}.
\end{itemize}

  The spectrum of a massless  Dirac fermion is shown in Figure~\ref{fig5}, which can be viewed as a limit of Figure~\ref{fig1} when $\mu\to 0$. Since $\mu =0$, the bound states take place at zero energy $\varepsilon =0$, so they are both critical and a supercritical points (other examples of zero energy confinement in different electric fields are given in \cite{ho14}). They cannot be considered as stable bound points since they represent a transition of resonance points from  positive to negative in the continuous spectrum, as we will see next.

   Resonance states for massive and massless Dirac particles, with different angular momenta, were also studied in detail. We define the concept of resonance wave functions by means of an eigenvalue problem with outgoing boundary conditions leading to complex eigenvalues, 
$\varepsilon= \varepsilon_R+ i \varepsilon_I$, and recall the meaning of the real and imaginary parts in terms of scattering states and Wigner delay time.
In Figure~\ref{fig3} a list of graphs is presented, for different heights of the potential well $v$, with eigenvalues of complex resonances and angular momentum $\ell=0, 2$, which give a complete image of the atomic collapse together with Figure~\ref{fig1} above on discrete energies (restricted to the interval $-\mu\leq \varepsilon\leq \mu$).
From this point of view, the bound states do not disappear, but sink into the continuous spectrum in the form of resonances. These plots display some important features: 
\begin{itemize}
\item[(i)]  The number of resonances (including bound states) is conserved, as a function of $ v $.

\item[(ii)]  All of them (at least those with a smaller imaginary part, which are the most physically important) flow in the same direction, from positive to negative $\varepsilon_R$.

\item[(iii)]   For higher angular momenta, the imaginary part $\varepsilon_I(v)$ is closer to the real axis, so these states are `more bounded'.

\end{itemize}
Analogous considerations can be applied to the massless case of Figure~\ref{fig7}. We have a special situation since for $\mu =0$, the bound states necessarily  take place at zero energy $\varepsilon =0$, which is both a critical and a supercritical point (see other examples of electric fields in \cite{ho14}). 
If we study the evolution of complex resonances $\varepsilon(v) =\varepsilon_R(v) + i \varepsilon_I(v)$ as we did with massive particles, we observe that they are very sensitive to the value of the angular momentum.
For zero angular momentum ($\ell=0$) the resonances are very weak (in fact, as we saw before, the critical wave function is not a good bound state).
 However, the behavior at $\ell=0$ is a relativistic effect, because in non-relativistic quantum mechanics, $\ell=0$ gives no centrifugal term.
It should be noted that a similar figure for the massless case appears in \cite{peeters08} (also in \cite{apalkov08} for $\mu=0$ and some values of angular momentum), but instead of complex values a broadening is given of the real values depending on the imaginary part. In none of the works on this problem have we found graphs like Figure~\ref{fig3} or similar in the massive case. 

 To better understand the resonances, we have calculated the phase shifts and the Wigner delay time of scattering states.
For a massive case we plot Figure~\ref{fig2} and for a massless case Figure~\ref{fig6}, showing that the phase change is large for depth $v$ near a critical point, and that there is a jump of $\pi$ at the origin for values of $v$ after capturing a bound energy level. 
This is in agreement with a relativistic version of Levinson's theorem.
The Wigner time delays are shown in Figure~\ref{fig4} (massive case) and in Figure~\ref{fig8} (massless case). In both figures the concordance of the complex eigenvalues of the outgoing states and the highest values of time delays is shown. There are some minor differences for the $\ell=0$ cases due to the special half-bound critical state.
All these results show consistency between resonances, critical values of collapses and time.

Most of the properties that have been shown in this work are based on numerical calculations on sharp potentials. In the near future, we plan to delve into these properties but especially using smoother potentials and analytical models.

 \section*{Acknowledgments}

 This study was supported by MCIN with funding from European Union NextGenerationEU (PRTR-C17.I1) and Consejeria de Educacion from JCyL through QCAYLE project, as well as PID2020-113406GB-I0 project by MCIN of Spain.
\c{S}.~K. thanks Ankara University and the warm hospitality of the Department of Theoretical Physics of the University of Valladolid, where this work has been carried out. L. S. appreciates the support of CONICET to develop this work at the University of Va\-lla\-dolid.

\end{document}